\documentclass[twocolumn,showpacs,preprintnumbers,amsmath,amssymb,nofootinbib]{revtex4-2}

\usepackage{bm}          
\usepackage{mathtools}       
\usepackage{comment} 
\usepackage{cancel}
\usepackage{graphicx}
\usepackage{dcolumn}
\usepackage[compact]{titlesec}
\usepackage{float}

\makeatletter
\DeclareRobustCommand{\cev}[1]{%
  {\mathpalette\do@cev{#1}}%
}
\newcommand{\do@cev}[2]{%
  \vbox{\offinterlineskip
    \sbox\z@{$\m@th#1 x$}%
    \ialign{##\cr
      \hidewidth\reflectbox{$\m@th#1\vec{}\mkern4mu$}\hidewidth\cr
      \noalign{\kern-\ht\z@}
      $\m@th#1#2$\cr
    }%
  }%
}
\makeatother
\usepackage{xcolor}

\definecolor{darkbrown}{RGB}{100, 20, 10}

\usepackage{hyperref}
\renewcommand{\arraystretch}{2.3}       
\newcommand{\half}{{{\textstyle\frac{1}{2}}}}
\newcommand{\quarter}{{{\textstyle\frac{1}{4}}}}
\newcommand{\be}{\begin{equation}}
\newcommand{\ee}{\end{equation} }
\newcommand{\beqa}{\begin{eqnarray} }
\newcommand{\eeqa}{\end{eqnarray} }
\newcommand{\ba}{\begin{array}}
\newcommand{\ea}{\end{array}}
\newcommand{\bpm}{\begin{pmatrix}}
\newcommand{\epm}{\end{pmatrix}}

\newcommand{\fR}{\mathfrak{R}}

\newcommand{\Deltab}{\mathbf{\Delta}}
\newcommand{\brDeltab}{\mathbf{\brDelta}}
\newcommand{\bBox}{\scalebox{1.1}{$\boldsymbol{\Box}$}}

\newcommand{\Spin}{\mathbf{Spin}}

\newcommand{\dis}{\displaystyle}

\newcommand{\Det}{{{\rm{Det\,}}}}

\newcommand{\rmd}{{\rm d}}

\newcommand\hcL{{\hat{\cal L}}}

\newcommand{\DO}{\mathbf{\nabla}}

\newcommand{\fT}{\mathfrak{T}}

\newcommand{\ODD}{\mathbf{O}(D,D)}

\newcommand\So{S_{\scriptscriptstyle{{(0)}}}}

\newcommand\Tr{{\scalebox{0.9}{${\mathrm{Tr}}$}}}

\newcommand\cA{{\cal A}}

\newcommand\cD{{\cal D}}

\newcommand\cF{{\cal F}}

\newcommand\cH{{\cal H}}

\newcommand\cJ{{\cal J}}

\newcommand\cO{{\cal O}}
\newcommand\cP{{\cal P}}

\def\tx{\tilde{x}}

\newcommand{\bs}{\scalebox{0.79}{$s$}}

\def\na{\nabla}

\def\bre{\bar{e}}

\def\breta{\bar{\eta}}

\def\brgamma{\bar{\gamma}}

\def\brp{{\bar{p}}}
\def\brq{{\bar{q}}}
\def\brr{{\bar{r}}}
\def\brs{{\bar{s}}}

\def\brDelta{{{\bar{\Delta}}}}

\def\brC{\bar{C}}

\def\brP{\bar{P}}

\def\brV{\bar{V}}

\def\brcF{\bar{\cF}}

\def\brcP{\bar{\cP}}

\newcommand\hR{{\hat{R}}}

\def\B{\mathtt{B}}

\newcommand{\trd}{{\bigtriangledown}}

\begin{document}


\title{Universal Box Operator: $\mathbf{O}(D,D)$-Symmetry and $\alpha^{\prime}$-Corrections}

\author{Kawon Lee}
\author{Jeong-Hyuck Park}
\affiliation{Department of Physics \& Center for Quantum Spacetime, Sogang 
University, Seoul 04107,  Korea}

\begin{abstract}
\centering\begin{minipage}{\dimexpr\paperwidth-6.9cm}
\noindent 
We construct a fully covariant, $\ODD$-symmetric d’Alembertian---or box operator---that acts on tensor fields of arbitrary rank and provides a universal kinetic term for all bosonic closed-string states. In its Riemannian parametrization, the operator packages the Riemann curvature, $H$-flux, and dilaton gradient into a single duality-covariant object. This yields $\ODD$-symmetric gravitational-wave equations for the massless sector, governs the tachyon and all massive modes, and clarifies how higher excitations contribute to  $\alpha^{\prime}$-corrections. The box operator thus supplies a unified description of closed-string dynamics across the entire spectrum. Our analysis shows that any apparent breaking of $\ODD$ symmetry arises only after integrating out massive modes in a Wilsonian sense, where loop momentum integrals obscure half of the doubled momenta. We stand on the view that $\ODD$ symmetry and doubled diffeomorphisms remain exact and undeformed at the fundamental level of string theory.
\end{minipage}  
\end{abstract}

\maketitle

Higher-derivative corrections encode genuinely stringy deviations from point-particle physics: they supplement familiar two-derivative low-energy actions with an infinite tower of higher-curvature terms~\cite{Fradkin:1984pq,Zwiebach:1985uq,Fradkin:1985fq,Callan:1985ia,Gross:1986mw,Cai:1986sa,Metsaev:1987zx,Hull:1987yi,Jack:1987mb,Bergshoeff:1988nn,Bergshoeff:1989de,Osborn:1989bu,deRoo:1992zp,Chemissany:2007he}, where the Regge slope $\alpha^{\prime}$ sets the fundamental length-squared scale. Because the metric $g_{\mu\nu}$ joins the skew-symmetric $B$-field and the dilaton $\phi$ to form the closed-string massless multiplet reshuffled by T-duality~\cite{Buscher:1987sk,Buscher:1987qj}, these higher-curvature terms are expected to respect the underlying $\mathbf{O}(D,D)$ symmetry. Achieving such a duality-covariant description remains an open challenge~\cite{Meissner:1996sa,Liu:2013dna,Godazgar:2013bja,Hohm:2015doa}: conventional Riemannian geometry supplies no duality-covariant curvature, and ad hoc insertions of the three-form flux ${H = \rmd B}$ obscure the symmetry they seek to preserve.

Double Field Theory (DFT)~\cite{Siegel:1993xq,Siegel:1993th,Hull:2009mi,Hull:2009zb,Hohm:2010jy} casts the two‑derivative effective actions into a manifestly 
$\ODD$‑covariant form~\cite{Hohm:2010pp,Hohm:2011ex,Hohm:2011zr,Jeon:2011sq,Grana:2012rr,Jeon:2012hp}. Here the term “double” refers to Duff’s doubled 
coordinates, ${x^{A} = (\tx_{\mu}, x^{\nu})}$~\cite{Duff:1989tf}: physical 
$D$‑dimensional dynamics emerges once half of the coordinates are gauged 
away~\cite{Park:2013mpa,Lee:2013hma}, or equivalently when the section condition 
${\partial_{A}\partial^{A}=\tilde{\partial}^{\mu}\partial_{\mu}+\partial_{\mu}\tilde{\partial}^{\mu}=0}$ is imposed. Replacing ordinary Riemannian geometry 
with a doubled differential calculus endowed with its own Ricci and scalar 
curvatures~\cite{Jeon:2010rw,Jeon:2011cn}, DFT unifies type IIA and type IIB 
supergravities—including all fermionic orders~\cite{Jeon:2012hp}—and yields 
Einstein‑like field equations intrinsic to the doubled 
spacetime~\cite{Angus:2018mep}. Yet a fully covariant counterpart of the 
four‑index Riemann tensor remains to be 
\textit{constructed}~\cite{Jeon:2011cn,Hohm:2011si}, preventing a manifestly 
duality‑invariant treatment of $\alpha^{\prime}$-corrections. 
Proposals that deform gauge transformations and algebras~\cite{Hohm:2013jaa,Bedoya:2014pma,Hohm:2014eba,Hohm:2014xsa,Lee:2015kba,Marques:2015vua,Hohm:2015mka,Lescano:2016grn,Baron:2017dvb,Baron:2018lve,Eloy:2020dko,Lescano:2021guc,Lescano:2024lwn,Hassler:2024yis,Gitsis:2024gfb,Lescano:2025yio} encounter potential 
obstructions at higher orders, and some skepticism remains regarding DFT’s 
ultimate relevance to full string theory~\cite{Hronek:2020xxi,Hsia:2024kpi}.

Considering the all-order string-field-theory results by Sen in a cosmological setting~\cite{Sen:1991zi}, supported by subsequent analyses~\cite{Hohm:2014sxa,Codina:2020kvj}, and the very existence of $\ODD$-symmetric string actions---bosonic~\cite{Lee:2013hma} and $\kappa$-symmetric superstring~\cite{Park:2016sbw}---which are fully covariant under both world-sheet and doubled target-spacetime diffeomorphisms in critical dimensions ${D=26}$ or ${10}$~\cite{Park:2020ixf}, we are inclined to believe that the $\ODD$ symmetry remains \textit{a priori} undeformed beyond the massless level.   {In our view, any apparent breaking of duality symmetry arises only after integrating out massive modes in a \textit{Wilsonian sense}: loop-momentum integrals constrained by the section condition involve only half of the doubled momenta and thus break manifest $\ODD$ covariance. 
Recent higher-order analyses~\cite{Hronek:2020xxi,Hsia:2024kpi}  seem to have indeed pointed out possible obstructions to maintaining manifest $\ODD$ covariance in  the higher $\alpha^{\prime}$-expansion of the massless effective action.
Our standpoint is complementary: such obstructions reflect the low-energy \textit{Wilsonian coarse-graining} rather than a fundamental deformation of the symmetry.}

In this Letter we close the curvature gap, clarify the mechanism behind 
$\ODD$‑symmetry breaking in $\alpha^{\prime}$-corrections, and simultaneously 
resolve a long‑standing coupling problem: how to describe, in a single 
duality‑covariant equation, the interaction between the closed-string massless 
sector and \emph{every} excitation of the string spectrum. For a bosonic 
closed string in flat Minkowski space, the well‑known mass-shell condition,
\be
\eta^{\mu\nu}p_{\mu}p_{\nu}+M^{2}=0\,,\quad\quad 
M^2=\frac{4}{\alpha^{\prime}}\big(N-1\big)\,,
\label{psquare}
\ee
with level number $N$ (${N=0}$ tachyon, ${N=1}$ massless sector, and ${N\ge2}$ 
massive modes), translates upon quantization into the Klein--Gordon equation, 
where $\eta^{\mu\nu}p_{\mu}p_{\nu}$ becomes the flat-spacetime d’Alembertian, 
$-\eta^{\mu\nu}\partial_{\mu}\partial_{\nu}$. The conventional minimal 
coupling would simply replace the flat metric $\eta^{\mu\nu}$ and the 
derivative $\partial_{\mu}$ by the curved metric $g^{\mu\nu}$ and the covariant 
derivative $\trd_{\mu}$, but unitarity of scattering amplitudes is known to 
require additional Riemann curvature terms~\cite{Porrati:1993in, 
Cucchieri:1994tx} (see also \cite{Christensen:1978md, Coimbra:2011nw, 
Coimbra:2014qaa, Cho:2015lha} for gravitational Dirac operators). The four‑index Riemann tensor captures the tidal effects on finite-sized or  highly excited  string 
modes~\cite{Giannakis:1998wi}, and T-duality then insists that it be 
accompanied by the $H$‑flux.

As the main result of this Letter, we introduce a pair of fully covariant, 
undeformed $\ODD$-symmetric d’Alembertians, $\Deltab$ and $\brDeltab$. These 
operators act on tensor fields of arbitrary rank and exhibit intriguing 
properties: their sum vanishes, $\Deltab+\brDeltab=0$, and their difference 
yields a \textit{box operator}, $\Deltab-\brDeltab=\bBox$, which augments the 
flat spacetime d’Alembertian $\eta^{\mu\nu}\partial_{\mu}\partial_{\nu}$. 
The full covariance of these operators means covariance under global 
$\ODD$ rotations and compatibility with the generalized Lie derivative 
of DFT~\cite{Siegel:1993th,Hull:2009zb}. 
Although constructed in an $\ODD$-symmetric minimal fashion, the operator becomes remarkably 
rich when the DFT variables are resolved into the trio $\{g,B,\phi\}$ through 
parametrization: $\bBox$ contains the Riemannian d’Alembertian 
$g^{\mu\nu}\trd_{\mu}\trd_{\nu}$ plus contractions with 
$R^{\kappa}{}_{\lambda\mu\nu}$, $H_{\lambda\mu\nu}$, and $\partial_{\mu}\phi$, 
thereby packaging  curvature, $H$-flux, and dilaton gradient in a single duality‑covariant object 
and reproducing the curvature coupling advocated in 
\cite{Porrati:1993in,Cucchieri:1994tx,Giannakis:1998wi}. 
It supplies the universal kinetic term valid for \textit{all} string modes and 
extends the flat spacetime mass-shell condition~(\ref{psquare}) to an 
arbitrary closed-string background:
\be
\left[\,\bBox\,-\frac{4}{\alpha^{\prime}}\big(N-1\big)\right]\Psi_{N}=0\,.
\label{KGeq}
\ee
Because the construction involves only genuine DFT fields, it extends unchanged 
to non‑Riemannian backgrounds~\cite{Morand:2017fnv,Cho:2019ofr} relevant for 
non‑relativistic strings and Newton–Cartan geometry~\cite{Gomis:2000bd,Danielsson:2000gi,Gomis:2005pg,Christensen:2013lma,Hartong:2015zia,Harmark:2017rpg,Harmark:2018cdl,Bergshoeff:2018yvt,Bergshoeff:2019pij,Harmark:2019upf,Bergshoeff:2021bmc,Oling:2022fft,Hartong:2022lsy}.    

Serving as a universal kinetic term, $\bBox$ governs the $\ODD$-consistent interaction of every string excitation with the closed-string massless sector. This formulation enables the one-loop integration of massive modes to be carried out explicitly and directly in the Wilsonian sense, yielding the associated $\alpha^{\prime}$-corrections. At the same time, it clarifies why the required momentum integrals generically break $\ODD$ symmetry.\vspace{3pt}

\section*{$\ODD$-Symmetric Fully-Covariant Box Operator}

Following the formalism and notation summarized in a recent 
review~\cite{Park:2025ugx}, we first explicitly present $\Deltab$ and then 
provide an essential explanation of its construction: 
\begin{widetext}
\be
\ba{l}
\Deltab T_{A_{1}A_{2}\cdots A_{\bs}}:=
P^{BC}\na_{B}\na_{C}T_{A_{1}A_{2}\cdots A_{\bs}}
+\dis{\sum_{i=1}^{\,\bs}~2P_{A_{i}}{}^{C}P_{B}{}^{D}\Big(\fR_{[CD]}-
\half\Gamma^{EF}{}_{C}\Gamma_{EFD}-\Gamma^{E}{}_{CD}\na_{E}\Big)T_{A_{1}\cdots 
A_{i-1}}{}^{B}{}_{A_{i+1}\cdots A_{\bs}}}\\
\quad+\dis{\sum_{i<j}~2\Big(
P_{A_{i}}{}^{D}P_{B}{}^{E}\fR_{A_{j}CDE}
+P_{A_{j}}{}^{D}P_{C}{}^{E}\fR_{A_{i}BDE}
-2P_{A_{i}}{}^{D}P_{B}{}^{E}P_{A_{j}}{}^{F}P_{C}{}^{G}S_{DEFG}\Big)T_{A_{1}\cdots 
A_{i-1}}{}^{B}{}_{A_{i+1}\cdots A_{j-1}}{}^{C}{}_{A_{j+1}\cdots A_{\bs}}}\,. 
\ea
\label{DELTA}
\ee
\end{widetext}
{The capital letters $A,B,\cdots$ denote the $\ODD$ vector indices,  which can be 
lowered and raised by means of the $\ODD$-invariant metric:
\be
\cJ_{AB}=\left(\ba{cc}{{0}}&{{1}}\\{{1}}&{{0}}\ea\right)\,.
\ee
The DFT-metric, often called  the ``generalized metric'', is then defined as a symmetric $\ODD$ element satisfying 
\be
\cH_{AB}=\cH_{BA}\,,\qquad
\cH_{A}{}^{C}\cH_{B}{}^{D}\cJ_{CD}=\cJ_{AB}\,.
\ee
From $\cJ_{AB}$ and $\cH_{AB}$ one constructs a pair of symmetric projectors~\cite{Jeon:2010rw}:
\be
\ba{lll}
P_{AB}=P_{BA}=\frac{1}{2}\big(\cJ_{AB}+\cH_{AB}\big)\,,\quad&\quad
P_{A}{}^{B}P_{B}{}^{C}=P_{A}{}^{C}\,,\\
\brP_{AB}=\brP_{BA}=\frac{1}{2}\big(\cJ_{AB}-\cH_{AB}\big)\,,\quad&\quad
\brP_{A}{}^{B}\brP_{B}{}^{C}=\brP_{A}{}^{C}\,,
\ea
\label{PbrP2}
\ee
which  are mutually  orthogonal and complete:
\be
\ba{ll}
P_{A}{}^{B}\brP_{B}{}^{C}=0\,, \qquad&\quad
P_{AB}+\brP_{AB}=\cJ_{AB}\,.
\ea
\label{CO}
\ee}
Together with the DFT-dilaton $d$, which yields the DFT integral measure $e^{-2d}$ 
(a scalar density with unit weight), these projectors comprise 
the DFT-Christoffel connection~\cite{Jeon:2011cn}:
\be
\ba{l}
\scalebox{0.92}{$
\Gamma_{CAB}=2\big(P\partial_{C}P\brP\big)_{[AB]\!}
+2\big(\brP_{[A}{}^{D}\brP_{B]}{}^{E}-
P_{[A}{}^{D}P_{B]}{}^{E}\big)\partial_{D}P_{EC}$}\\
\scalebox{1.05}{$\,-4\left(\frac{P_{C[A}P_{B]}{}^{D}}{P_{M}{}^{M}-
1}+\frac{\brP_{C[A}\brP_{B]}{}^{D}}{\brP_{M}{}^{M}-
1}\right)\!\Big(\partial_{D}d+(P\partial^{E}P\brP)_{[ED]}\Big)\,,$}
\ea
\label{Gamma}
\ee
which satisfies the following torsionless and symmetric properties:
\be
\ba{ll}
\Gamma_{ABC}+\Gamma_{BCA}+\Gamma_{CAB}=0\,,\quad&\quad
\Gamma_{CAB}+\Gamma_{CBA}=0\,.
\ea
\label{symG}
\ee
The semi-covariant derivative of a generic tensor density with weight $\omega$ 
is defined as follows:
\be
\ba{rll}
\na_{C}T_{A_{1}A_{2}\cdots A_{\bs}}
&=&\partial_{C}T_{A_{1}A_{2}\cdots A_{\bs}}-\omega\Gamma^{B}{}_{BC}\, 
T_{A_{1}A_{2}\cdots A_{\bs}}\\
{}&{}&+
\dis{\sum_{i=1}^{\,\bs}}~\Gamma_{CA_{i}}{}^{B}\, T_{A_{1}\cdots 
A_{i-1}BA_{i+1}\cdots A_{\bs}}\,.
\ea
\label{semicovD}
\ee
The concept of “semi-covariance” serves as an intermediate step in the 
formalism, facilitating the construction of fully covariant derivatives and 
curvatures. By appropriately projecting the derivative and tensor indices, full 
covariance can be achieved~\cite{Jeon:2011cn}. For example:
\be
\ba{l}
P_{C}{}^{D}{\brP}_{A_{1}}{}^{B_{1}}{\brP}_{A_{2}}{}^{B_{2}}\cdots{\brP}_{A_{\bs}}{}^{B_{\bs}}
\DO_{D}T_{B_{1}B_{2}\cdots B_{\bs}}\,,\\
P^{CD}{\brP}_{A_{2}}{}^{B_{2}}{\brP}_{A_{3}}{}^{B_{3}}\cdots{\brP}_{A_{\bs}}{}^{
B_{\bs}}
\DO_{C}T_{DB_{2}B_{3}\cdots B_{\bs}}\,.
\ea
\label{covD}
\ee
Even so, the compatibility condition holds automatically for the fundamental 
fields of DFT with full covariance:
\be
\ba{ll}
\na_{C}P_{AB}=0= \na_{C}\brP_{AB}\,,\quad&\quad\!\!
\na_{C}d=-\half e^{2d}\na_{C}\big(e^{-2d}\big)=0\,.
\ea
\label{compatibility}
\ee
The second line in (\ref{DELTA}) contains the “field strength” of the DFT-
Christoffel connection, which is not covariant:
\be
\scalebox{0.91}{$\fR^{C}{}_{DAB}=\partial_{A}\Gamma_{B}{}^{C}{}_{D}-
\partial_{B}\Gamma_{A}{}^{C}{}_{D}+\Gamma_{A}{}^{C}{}_{E}\Gamma_{B}{}^{E}{}_{D}-
\Gamma_{B}{}^{C}{}_{E}\Gamma_{A}{}^{E}{}_{D}\,,$}
\label{FSR}
\ee 
and the associated semi-covariant four-index curvature:
\be
S_{ABCD}:=\half\left(\fR_{ABCD}+\fR_{CDAB}-
\Gamma^{E}{}_{AB}\Gamma_{ECD}\right)\,.
\label{RiemannS}
\ee
These satisfy several important identities~\cite{Park:2025ugx}, including:
\be
\ba{ll}
\fR_{ABCD}=\fR_{[AB][CD]}\,,\quad&\quad S_{ABCD}=S_{[AB][CD]}=S_{CDAB}\,,
\label{fRSsym}
\ea
\ee
and yield 
\be
\ba{ll}
\fR_{AB}=\fR^{C}{}_{ACB}\neq \fR_{BA}\,,\quad&\qquad 
S_{AB}=S_{BA}=S^{C}{}_{ACB}\,.
\ea
\ee
The $\ODD$-symmetric fully-covariant DFT versions of the scalar and two-index 
Ricci curvatures are then~\cite{Jeon:2011cn}:
\be
\ba{rll}
\So&=&(P^{AC}P^{BD}-\brP^{AC}\brP^{BD})S_{ABCD}\,,\\
(PS\brP)_{AB}&=&P_{A}{}^{C}\brP_{B}{}^{D}S_{CD}\,.
\ea
\label{SoR}
\ee

Now, by replacing every explicit occurrence of the projector $P_{AB}$ in the 
definition of $\Deltab$~(\ref{DELTA}) with the opposite projector $\brP_{AB}$, 
we define the mirroring operator:
\be
\brDeltab T_{A_{1}A_{2}\cdots A_{\bs}}:=
\brP^{BC}\na_{B}\na_{C}T_{A_{1}\cdots 
A_{\bs}}\,+\sum_{i=1}^{\,\bs}~\cdots~+\sum_{i<j}~\cdots\,.
\label{brDELTA}
\ee
See Supplementary  Material (SM) for the full expression. As shown in  SM, their sum vanishes identically, thanks to 
the section condition:
\be
\Deltab+\brDeltab=\cJ^{AB}\na_{A}\na_{B}+~\cdots~ =0\,.
\label{SUM}
\ee
This can be viewed as the covariant version of the section condition 
(cf.~\cite{Cederwall:2014kxa}). Their difference defines the fully-
covariant box operator:
\be
\bBox:=\Deltab-\brDeltab=\cH^{AB}\na_{A}\na_{B}+~\cdots~\,.
\label{BOXOP}
\ee

While we refer to SM for the direct verification of the full covariance of 
$\Deltab$ and $\brDeltab$ (and consequently $\bBox$), we emphasize that no 
individual component within $\Deltab$~(\ref{DELTA}) or 
$\brDeltab$~(\ref{brDELTA}) is fully covariant on its own. Only the complete 
expressions achieve full covariance, and the inclusion of the four-index 
curvatures is essential for this.\vspace{3pt}

\section*{Applications}

\subsection{Tachyon at Level Zero: ${N=0}$}
We first consider a scalar field (the tachyon). With 
$\Gamma^{A}{}_{AB}=-2\partial_{B}d$, we reproduce a result of 
\cite{Jeon:2011cn}:
\be
\bBox\Phi=\cH^{AB}\na_{A}\na_{B}\Phi=e^{2d}\partial_{A}\left(e^{-
2d}\cH^{AB}\partial_{B}\Phi\right)\,,
\ee
which is consistent with an $\ODD$-symmetric action:
\be
\!-\int_{\Sigma_{D}}\!\! e^{-
2d}\big(\cH^{AB}\partial_{A}\Phi\partial_{B}\Phi+M^2\Phi^2\big)\!=\!\int_{\Sigma
_{D}}\!\!e^{-2d}\Phi(\bBox-M^2)\Phi\,,
\ee
where $\Sigma_{D}$ is a $D$-dimensional integral domain (section). 

Solving the section condition by letting ${\tilde{\partial}^{\mu}\equiv 0}$ and 
assuming the well-known Riemannian parametrization~\cite{Giveon:1988tt}:
\be
\ba{ll}
\cH_{AB}=\!\left(\ba{cc}
g^{-1}&-g^{-1}B\\
Bg^{-1}&
~g-Bg^{-1}B\ea\right)\,,~~~& e^{-2d}=\sqrt{-g}\,e^{-2\phi}\,,
\ea
\label{RP}
\ee
we get, with the conventional covariant derivative $\trd_{\mu}$,
\be
\bBox\Phi=e^{2\phi}\trd_{\mu}\left(e^{-2\phi}\partial^{\mu}\Phi\right)\,.
\label{BoxPhi}
\ee

\subsection{Massless Sector: Gravitational Wave Equations}
The Einstein Double Field Equations~\cite{Angus:2018mep},
\be
{G_{AB}=T_{AB}}\,,
\ee
serve as the unified equations of motion for the closed-string massless sector. 
Here, $G_{AB}$ and $T_{AB}$ represent the off-shell and on-shell conserved 
$\ODD$-symmetric DFT versions of the Einstein curvature tensor~\cite{Park:2015bza} 
and the energy-momentum tensor, respectively. In particular, with (\ref{SoR}),
\be
G_{AB}=4(PS\brP)_{[AB]}-\half\cJ_{AB}\So~~:~~\na_{A}G^{A}{}_{B}=0\,.
\ee
It is worth noting that $G_{AB}$ allows separate decompositions into the 
scalar and Ricci curvatures~(\ref{SoR}):
\be
\ba{ll}   
G_{A}{}^{A}=-D\So\,,\qquad&\qquad (PG\brP)_{AB}=2(PS\brP)_{AB}\,.
\ea
\ee
We proceed to linearize the curvatures (cf.~\cite{Ko:2015rha}):
\be
\ba{rll}
\delta \So&=&2\bBox\delta d-\na_{A}\Omega^{A}+2(PS\brP)^{AB}\delta\cH_{AB}\,,\\
\delta (PS\brP)_{AB}&=&-
\quarter\bBox(P\delta\cH\brP)_{AB}+P_{A}{}^{[C}\brP_{B}{}^{D]}\na_{C}\Omega_{D}\,\\
{}&{}&-\half(PS\brP\delta \cH P)_{AB}+\half (\brP\delta\cH PS\brP)_{AB}\,,
\ea
\ee
where we have introduced a shorthand notation:
\be
\Omega_{A}=e^{2d}\na_{B}\delta\!\left(e^{-
2d}\cH^{B}{}_{A}\right)=\na_{B}\delta\cH^{B}{}_{A}-
2\cH_{A}{}^{B}\partial_{B}\delta d\,.
\label{OmegaHG}
\ee
This combination is fully covariant, given (\ref{covD}) and the identity:
\be
\delta\cH_{AB}=(P\delta\cH\brP)_{AB}+(\brP\delta\cH P)_{AB}\,.
\ee
Consequently, the linearized Einstein Double Equations around a generic 
vacuum (where ${T_{AB}=G_{AB}=0}$), represented by ${\delta G_{AB}\equiv 0}$, 
lead to $\ODD$-symmetric gravitational wave equations governed by the box 
operator:
\be
\ba{ll}
\bBox(P\delta\cH\brP)_{AB}\equiv 0\,,\qquad~\qquad\bBox\delta d\equiv 0\,,
\ea
\label{GWAVE}
\ee
provided that the $\ODD$-symmetric harmonic gauge condition, 
${\Omega_{A}\equiv 0}$, is imposed. This nontrivial result demonstrates the 
relevance of the box operator to string theory, particularly highlighting its 
connection to the mass-shell conditions~(\ref{psquare}), (\ref{KGeq}). 

The projected $\ODD$ indices can be represented by $D$-dimensional vector 
indices associated with the twofold spin groups, 
$\Spin(1,{D-1})\times\Spin({D-1},1)$. These indices are denoted by unbarred 
small letters $p,q,\cdots$, or barred ones $\brp,\brq,\cdots$, which conform 
to the Minkowskian flat metrics $\eta_{pq}$ or $\breta_{\brp\brq}$, 
respectively. This representation is achieved through contraction with the DFT-vielbeins $\{V_{Ap}, \brV_{A\brp}\}$, which square to the 
projectors~\cite{Jeon:2011cn,Jeon:2011vx,Park:2025ugx}: 
\be
\ba{cc}
V_{A}{}^{p}V_{B}{}^{q}\eta_{pq}=P_{AB}\,,\qquad&\qquad
\brV_{A}{}^{\brp}\brV_{B}{}^{\brq}\breta_{\brp\brq}=\brP_{AB}\,,
\ea
\ee
and for a mixed-index tensor one finds:
\be
(PT\brP)_{AB}=V_{A}{}^{p}\brV_{B}{}^{\brq}T_{p\brq}\quad\Longleftrightarrow\quad
T_{p\brq}=T_{AB}V^{A}{}_{p}\brV^{B}{}_{\brq}\,.
\ee
The semi-covariant derivative $\na_{A}$~(\ref{semicovD}) generalizes to the 
master semi-covariant derivative $\cD_{A}$~\cite{Jeon:2011vx, Jeon:2011sq}. By 
incorporating not only the Christoffel connection~(\ref{Gamma}) but also two 
sets of spin connections, the derivative is fully compatible with the DFT-vielbeins: $\cD_{A}V_{Bq}=0=\cD_{A}\brV_{B\brq}$. Consequently, it enables us 
to transform $\Deltab (PT\brP)_{AB}$ into a more compact expression 
(cf.~(\ref{DeltabCOMPACT}) for $\brDeltab T_{p\brq}$):
\be
\ba{rll}\vspace{-3pt}
\Deltab T_{p\brq}
&=&\cD_{r}\cD^{r}T_{p\brq}+2\fR_{\brq \brs pr}T^{r\brs}\\
{}&{}&+2\big(\fR_{[pr]}-\half\Gamma^{AB}{}_{p}\Gamma_{ABr}-
\Gamma^{C}{}_{pr}\cD_{C}\big)T^{r}{}_{\brq}\,.
\ea
\label{DeltabCOMPACT}
\ee

The Riemannian background of (\ref{RP}) yields~\cite{Jeon:2011vx}
\be
\!\!\ba{ll}
V_{Ap}\!=\!\frac{1}{\sqrt{2}}\!\!\left(\!\!\ba{c}\vspace{-5pt}e_{p}{}^{\mu}\\
e_{\nu}{}^{q}\eta_{qp}{+B_{\nu\sigma}}e_{p}{}^{\sigma}\ea\!\!\!\right)\!,&\!\!
\brV_{A\brp}\!=\!\frac{1}{\sqrt{2}}\!\!\left(\!\!\ba{c}\vspace{-
5pt}\bre_{\brp}{}^{\mu}\\
\bre_{\nu}{}^{\brq}\breta_{\brq\brp}{+B_{\nu\sigma}}\bre_{\brp}{}^{\sigma}\ea\!\!
\!\right)\!,
\ea
\label{00V}
\ee
where $e_{\mu}{}^{p}$ and $\bre_{\mu}{}^{\brp}$ are ordinary vielbeins which 
square to the same metric: $e_{\mu}{}^{p}e_{\nu}{}^{q}\eta_{pq}=-
\bre_{\mu}{}^{\brp}\bre_{\nu}{}^{\brq}\breta_{\brp\brq}=g_{\mu\nu}\,$. 

By letting ${\fT_{\mu\nu}=e_{\mu}{}^{p}\bre_{\nu}{}^{\brq} T_{p\brq}}$, the box 
operator leads to 
\be
\ba{rll}
e_{\mu}{}^{p}\bre_{\nu}{}^{\brq}\bBox T_{p\brq}&=& e^{2\phi}\trd^{\rho}\big(e^{-
2\phi}\trd_{\rho}\fT_{\mu\nu}\big)+2\hR_{\mu}{}^{\rho}{}_{\nu}{}^{\sigma}
\fT_{\rho\sigma}\\
{}&{}&- H_{\rho\sigma\mu} \trd^{\rho}\fT^{\sigma}{}_{\nu}+
H_{\rho\sigma\nu}\trd^{\rho}\fT_{\mu}{}^{\sigma}\,,
\ea
\ee
where the Riemann curvature tensor appears through
\be
\ba{rll}
\hR_{\mu}{}^{\rho}{}_{\nu}{}^{\sigma}&=&R_{\mu}{}^{\rho}{}_{\nu}{}^{\sigma}
-\half H_{(\mu}{}^{\rho\kappa}H_{\nu)}{}^{\sigma}{}_{\kappa}-\quarter  
H_{\mu\nu\kappa}H^{\rho\sigma\kappa}\\
{}&{}&
+\half\trd_{(\mu}H_{\nu)}{}^{\rho\sigma}+\half\trd^{(\rho}H^{\sigma)}{}_{\mu\nu}
\\
{}&{}&
+\frac{1}{4}\delta_{\mu}{}^{\rho}\Big[e^{2\phi}\trd_{\kappa}\big(e^{-
2\phi}H^{\kappa\sigma}{}_{\nu}\big)-
\frac{1}{2}H_{\nu\kappa\lambda}H^{\sigma\kappa\lambda}\Big]\\
{}&{}&
-\frac{1}{4}\Big[e^{2\phi}\trd_{\kappa}\big(e^{-
2\phi}H^{\kappa\rho}{}_{\mu}\big)+\frac{1}{2}H_{\mu\kappa\lambda}H^{\rho\kappa\lambda}\Big]\delta_{\nu}{}^{\sigma}
\,.
\ea
\label{RiemannHflux}
\ee
For the stringy gravitational wave equations (\ref{GWAVE}), $\fT_{\mu\nu}$ is 
identified as $e_{\mu}{}^{p}\bre_{\nu}{}^{\brq}\delta\cH_{p\brq}=\delta 
g_{\mu\nu}-\delta B_{\mu\nu}$, while $\delta d=\delta \phi-\quarter 
g^{\mu\nu}\delta g_{\mu\nu}$.\vspace{3pt}

\subsection{Massive Modes: Higher-Order $\alpha^{\prime}$-Corrections}
The box operator naturally sets the kinetic term for each string mode, up to 
possible gauge fixing:
\be
\int_{\Sigma_{D}}T_{A_{1}A_{2}\cdots 
A_{\bs}}\left[\bBox-\frac{4}{\alpha^{\prime}}(N{-1})\right]T^{A_{1}A_{2}\cdots 
A_{\bs}}\,,
\ee
where the tensor indices are generically projected (due to the level matching 
condition, e.g. $T_{AB}=V_{A}{}^{p}\brV_{B}{}^{\brq}T_{p\brq}$). Here we have set $\omega=\frac{1}{2}$ for the weight for simplicity. 

Integrating out the massive string modes, we obtain the one-loop determinant 
for the $\alpha^{\prime}$-corrections: 
\be
-\frac{1}{2}\ln\Det\!\left[1-\frac{\alpha^{\prime}\bBox}{4(N-
1)}\right]=\sum_{n=1}^{\infty}~\frac{1}{2n}\Tr\left(\frac{\alpha^{\prime}\bBox}{
4(N-1)}\right)^{\!n}\,.
\label{lnDet}
\ee
Evaluating the trace involves summing over the tensor indices and performing a momentum integral. By substituting $\partial_{A}$ with a $c$-number-valued (diagonal) operator $ik_{A}$, the box operator~(\ref{BOXOP}) 
schematically takes the following quadratic form:
\be
\bBox= -\cH^{AB}k_{A}k_{B}+\half\cO^{A}k_{A}+\half k_{A}\cO^{A}+\cO[\fR]\,,
\label{bBoxk}
\ee
where $\cO^{A}$ and $\cO[\fR]$ are $\ODD$-symmetric operators. Crucially, the 
section condition renders half of the components in $k_{A}$ trivial, meaning 
not all components of $\cH_{AB}$~(\ref{RP}) and $\cO_{A}=(\cO^{\mu},\widetilde{\cO}_{\nu})$ participate in the $\alpha^{\prime}$-corrections~(\ref{lnDet}). This results in the breaking of 
the $\ODD$ symmetry. Specifically, solving the section condition by 
setting ${\tilde{\partial}^{\mu}\equiv 0}$, we have 
$k_{A}=(\tilde{k}^{\mu},k_{\nu})\equiv(0,k_{\nu})$, and thus
\be
\bBox \equiv \quarter 
g_{\mu\nu}\cO^{\mu}\cO^{\nu}-g^{\mu\nu}\!\left(k_{\mu}{-\half} 
g_{\mu\rho}\cO^{\rho}\right)\!\left(k_{\nu}{-\half} 
g_{\nu\sigma}\cO^{\sigma}\right)+\cO[\fR]\,.
\ee
The subsequent momentum integral then produces $\ODD$-breaking terms, such as 
$\quarter g_{\mu\nu}\cO^{\mu}\cO^{\nu}$ rather than its $\ODD$-completion 
$\quarter\cH_{AB}\cO^{A}\cO^{B}$. This highlights how integrating out the 
massive modes leads to the breaking of the $\ODD$ symmetry in our approach. 

Although we have not explicitly summed all contributions from every 
massive mode, it is evident from (\ref{fRSsym}) that the Riemann curvature 
tensor begins to appear in the $\alpha^{\prime}$-corrections~(\ref{lnDet}) at 
quadratic order (starting from $n=2$), in agreement with 
\cite{Fradkin:1984pq}.

\section*{Conclusion}
We have constructed a fully covariant, $\ODD$-symmetric box operator 
\[
\bBox = \Deltab - \brDeltab \,,
\] 
which acts on arbitrary tensor fields and provides a universal kinetic term for the entire bosonic closed-string spectrum. In its Riemannian parametrization, the operator packages the Riemann curvature, $H$-flux, and dilaton gradient into a single duality-covariant object. This encompasses the tachyon (${N=0}$), the massless sector (${N=1}$)---where it leads to $\ODD$-symmetric gravitational-wave equations in harmonic gauge---and the massive modes (${N \geq 2}$), where it provides their universal kinetic terms in agreement with \cite{Porrati:1993in, Cucchieri:1994tx}.  

Our construction provides strong evidence that $\ODD$ symmetry and doubled diffeomorphisms are never broken nor deformed at the fundamental level. The same operator applies uniformly to all levels of the string spectrum. The \emph{appearance} of $\ODD$ breaking occurs only after integrating out massive excitations in a Wilsonian sense: loop momentum integrals under the section condition remove half of the doubled momenta, leaving only Riemannian completions and thus hiding the duality symmetry in the effective action. Hence, any reduction of symmetry is an effective artifact of coarse-graining, not a fundamental feature of string theory.  

This unified framework identifies undeformed $\ODD$ symmetry and doubled diffeomorphisms as the organizing principles for closed-string dynamics, while clarifying how their manifest form can be obscured in low-energy $\alpha^{\prime}$-corrections. It provides a novel foundation for finite-$\alpha^{\prime}$ physics---including black-hole thermodynamics~\cite{Wald:1993nt,Jacobson:1994qe,Sen:2005iz,Sahoo:2006pm,Cvitan:2007hu} and 
early-universe scenarios~\cite{Antoniadis:1993jc,Gasperini:1996fu,Gasperini:2002bn,Antoniadis:2024ypf}---and suggests natural extensions to the Ramond--Ramond sector via spinorial generalizations of the box operator~\cite{footnoteCON}. Although our derivation was carried out for the bosonic string, the construction applies verbatim to the bosonic subsector of superstring theory.  

{We firmly stand on the view that $\ODD$ symmetry and doubled diffeomorphisms remain exact and undeformed at the fundamental level of (super)string theory.}\\

\noindent{\textit{Acknowledgments.}}---We wish to thank Martin Cederwall, Junho 
Hong, Wontae Kim, and Dimitrios Tsimpis for helpful discussions and 
correspondence. This work is supported by the National Research Foundation of 
Korea (NRF) through grants RS-2023-NR077094 and RS-2020-NR049598 (Center 
for Quantum Spacetime: CQUeST).




\begin{widetext}
\appendix
\setcounter{equation}{0}
\setcounter{figure}{0}
\renewcommand{\theequation}{SM\,\arabic{equation}}
\renewcommand{\thefigure}{SM\,\arabic{figure}}
\setlength{\jot}{9pt}                 
\renewcommand{\arraystretch}{3.2}

\begin{center}
\vspace{23pt}
\vspace{23pt}
\vspace{24pt}
	\Large	\textbf{Supplementary Material (SM)}
\end{center}
\vspace{5pt}

\section*{SM-A: Notation and  Useful Relations~\cite{Park:2025ugx}}
The generalized  Lie derivative, denoted as $\hcL_{\xi}$,   is  explicitly given by~\cite{Siegel:1993th,Hull:2009zb}:
\be
\hcL_{\xi}T_{A_{1}\cdots A_{\bs}}=\xi^{B}\partial_{B}T_{A_{1}\cdots A_{\bs}}+\omega\partial_{B}\xi^{B}\,T_{A_{1}\cdots A_{\bs}}+\sum_{i=1}^{\,\bs}~(\partial_{A_{i}}\xi_{B}-\partial_{B}\xi_{A_{i}})\,T_{A_{1}\cdots A_{i-1}}{}^{B}{}_{A_{i+1}\cdots  A_{\bs}}\,,
\label{hcL}
\ee
where $\omega$ is the weight of the tensor density. Each  index $A_{i}$ is  infinitesimally rotated by an  $\mathbf{so}(D,D)$ element, $2\partial_{[A_{i}}\xi_{B]}$.

The  pair of two-index projectors~(\ref{PbrP2}), which are orthogonal and complete~(\ref{CO}), further yield a pair of six-index projectors~\cite{Jeon:2011cn}:
\be
\ba{ll}
\cP_{ABC}{}^{DEF}=P_{A}{}^{D}P_{[B}{}^{[E}P_{C]}{}^{F]}+\textstyle{\frac{2}{P_{M}{}^{M}-1}}P_{A[B}P_{C]}{}^{[E}P^{F]D}\,,&\quad
\bar{\cP}_{ABC}{}^{DEF}=\brP_{A}{}^{D}\brP_{[B}{}^{[E}\brP_{C]}{}^{F]}+\textstyle{\frac{2}{\brP_{M}{}^{M}-1}}\brP_{A[B}\brP_{C]}{}^{[E}\brP^{F]D}\,.
\ea
\label{P6}
\ee
These satisfy  the  projection properties:
\be
\ba{ll}
\cP_{ABC}{}^{DEF}\cP_{DEF}{}^{GHI}=\cP_{ABC}{}^{GHI}\,,\qquad&\qquad
\brcP_{ABC}{}^{DEF}\brcP_{DEF}{}^{GHI}=\brcP_{ABC}{}^{GHI}\,,
\ea
\ee
some symmetric  relations:
\be
\ba{ll}
\cP_{ABCDEF}=\cP_{DEFABC}=\cP_{A[BC]D[EF]}\,,
\qquad&\qquad \cP_{[AB]CDEF}=\cP_{CAB[EF]D}\,,\\
\brcP_{ABCDEF}=\brcP_{DEFABC}=\brcP_{A[BC]D[EF]}\,,
\qquad&\qquad \brcP_{[AB]CDEF}=\brcP_{CAB[EF]D}\,,
\ea
\label{symP6}
\ee
and traceless conditions:
\be
\ba{ll}
\quad P^{AB}\cP_{ABCDEF}=0\,,\qquad&\qquad
\quad \brP^{AB}\brcP_{ABCDEF}=0\,.
\ea
\label{symP6s}
\ee
The six-index projectors organize the anomalous terms of the semi-covariant quantities under   the  diffeomorphism  transformations set by the generalized Lie derivative~(\ref{hcL}), as demonstrated  in the following sections of SM.

It is useful to note from the explicit expression of $\Gamma_{CAB}$, or from the compatibility condition $\na_{C}d=0$~(\ref{compatibility}), that
\be
\ba{ll}
P_{A}{}^{D}\brP_{B}{}^{E}\Gamma_{CDE}=\big(P\partial_{C}P\brP\big)_{AB}\,,\qquad&\qquad \Gamma^{B}{}_{BC}=-2\partial_{C}d\,,
\ea
\label{divC}
\ee
where  the free index $C$ is derivative-index-valued, such that for example:
\be
\ba{rrr}
\Gamma^{C}{}_{AB}\partial_{C}=\big(P_{A}{}^{D}P_{B}{}^{E}+\brP_{A}{}^{D}\brP_{B}{}^{E}\big)\Gamma^{C}{}_{DE}\partial_{C}\,,\qquad&\qquad
P_{A}{}^{D}\brP_{B}{}^{E}\Gamma^{C}{}_{DE}\partial_{C}=0\,,\qquad&\qquad
\Gamma_{B}{}^{BC}\partial_{C}=0\,,\\
P_{A}{}^{E}\brP_{B}{}^{F}\Gamma^{N}{}_{EF}P_{C}{}^{G}\brP_{D}{}^{H}\Gamma_{NGH}=0\,,\qquad&\qquad
\partial_{E}\big(P_{A}{}^{C}\brP_{B}{}^{D}\Gamma^{E}{}_{CD}\big)=0\,,\qquad&\qquad \Gamma^{A}{}_{AC}\Gamma_{B}{}^{BC}=0\,.
\ea
\label{PbrPG}
\ee
The field strength of the Christoffel connection $\fR_{ABCD}$ and the semi-covariant four-index curvature $S_{ABCD}$ satisfy~\cite{Jeon:2011cn}
\be
\ba{ll}
\fR_{ABCD}=\fR_{[AB][CD]}=(P_{A}{}^{E}P_{B}{}^{F}+\brP_{A}{}^{E}\brP_{B}{}^{F})\fR_{EFCD}\,,\qquad&\qquad S_{ABCD}=S_{CDAB}=S_{[AB][CD]}\,,
\ea
\ee
with additional properties:
\be
\ba{lll}
S_{A[BCD]}=0\,,\qquad&~~~~~~\qquad
P_{A}{}^{E}P_{B}{}^{F}\brP_{C}{}^{G}\brP_{D}{}^{H}S_{EFGH}=0\,,\qquad&~~~~~\qquad
P_{A}{}^{E}\brP_{B}{}^{F}P_{C}{}^{G}\brP_{D}{}^{H}S_{EFGH}=0\,.
\ea
\label{RSp}
\ee

From the symmetric properties of the DFT-Christoffel connection, we deduce 
\be
\Gamma_{ACD}\Gamma^{CD}{}_{B}=\Gamma_{A[CD]}\Gamma^{[CD]}{}_{B}=-\frac{1}{2}\Gamma_{ACD}\Gamma_{B}{}^{CD}\,,
\label{GGGG}
\ee
which simplifies the  expression for $\fR_{[AB]}$: 
\be
\fR_{[AB]}=\fR_{C[A}{}^{C}{}_{B]}=-\half\partial_{C}\Gamma^{C}{}_{AB}-\half\Gamma_{C}{}^{CD}\Gamma_{DAB}\,.
\label{fRsAB}
\ee

Mirroring (\ref{DeltabCOMPACT}), we have the following compact expression for $\brDeltab (PT\brP)_{AB}$:
\be
\brDeltab T_{p\brq}=\brDeltab (PT\brP)_{AB}V^{A}{}_{p}\brV^{B}{}_{\brq}=
\cD_{\brr}\cD^{\brr}T_{p\brq}+2\fR_{pr\brq\brs}T^{r\brs}
+2\big(\fR_{[\brq\brs]}-\half\Gamma^{AB}{}_{\brq}\Gamma_{AB\brs}-\Gamma^{C}{}_{\brq\brs}\cD_{C}\big)T_{p}{}^{\brs}\,.
\label{brDeltabCOMPACT}
\ee

Without loss of generality, we can parametrize a generic two-index tensor $T_{AB}$ as follows:
\be
T_{AB}=\left[\scalebox{0.9}{$\left(\ba{cc}1&0\\B&1\ea\right)\left(\ba{cc} T_{(1,1)}&T_{(1,2)}\\T_{(2,1)}&T_{(2,2)}\ea\right)\left(\ba{cc}1&-B\\0&1\ea\right)$}\right]_{AB}\,,
\ee
and compute the following projection  for the Riemannian  DFT-background characterized by (\ref{RP}):
\be
(PT\brP)_{AB}=\frac{1}{4}\left[\scalebox{0.9}{$
\left(\ba{c}1\\ g+B\ea\right)
\Big(T_{(1,1)}+g^{-1}T_{(2,1)}-T_{(1,2)}g^{-1}-g^{-1}T_{(2,2)}g^{-1}\Big)\left(\ba{cc}1&~{-g-B}\ea\right)$}\right]_{AB}\,.
\ee
Consequently,  for $T_{p\brq}=e_{p}{}^{\mu}\bre_{\brq}{}^{\nu}\fT_{\mu\nu}$,  we note:
\be
\fT^{\mu\nu}=g^{\mu\rho}g^{\nu\sigma}\fT_{\rho\sigma}=-\frac{1}{2}\Big(T_{(1,1)}+g^{-1}T_{(2,1)}-T_{(1,2)}g^{-1}-g^{-1}T_{(2,2)}g^{-1}\Big)^{\mu\nu}\,.
\ee

The variation of the Riemannian DFT-metric~(\ref{RP})  is given by:
\be
\ba{ll}
\delta \cH_{AB}=\left[\scalebox{0.9}{$\left(\ba{cc}1&0\\B&1\ea\right)\!\left(\ba{cc} 
-g^{-1}\delta gg^{-1}~&~-g^{-1}\delta B\\\delta B g^{-1}~&~\delta g
\ea\right)\!\left(\ba{cc}1&-B\\0&1\ea\right)$}\right]_{AB},\quad&\quad
\delta\cH_{p\brq}=V^{A}{}_{p}\brV^{B}{}_{\brq}\delta\cH_{AB}=
e_{p}{}^{\mu}\bre_{\brq}{}^{\nu}(\delta g_{\mu\nu}-\delta B_{\mu\nu})\,,
\ea
\ee
such that the quantity for the $\ODD$-symmetric harmonic gauge, $\Omega_{A}$ in (\ref{OmegaHG}), assumes the form:
\be
\Omega_{A}=
\na_{B}\delta\cH^{B}{}_{A}-2\cH_{A}{}^{B}\partial_{B}\delta d=-\left(C^{\mu}\,,\,\brC_{\nu}+B_{\nu\rho}C^{\rho}\right)\,,
\ee
where
\be
\ba{ll}
C_{\mu}=g_{\mu\nu}C^{\nu}=e^{2\phi}\trd^{\rho}\!\left(e^{-2\phi}\delta g_{\rho\mu}\right)
-\frac{1}{2}H_{\mu}{}^{\rho\sigma}\delta B_{\rho\sigma}+2\partial_{\mu}\big(\delta\phi-\quarter g^{\rho\sigma}\delta g_{\rho\sigma}\big)\,,\qquad&\qquad
\brC_{\mu}=e^{2\phi}\trd^{\rho}\big(e^{-2\phi}\delta B_{\rho\mu}\big)\,.
\ea
\label{Cs}
\ee

Regarding $\hR_{\mu}{}^{\rho}{}_{\nu}{}^{\sigma}$~(\ref{RiemannHflux}), it is worth while to note:
\be
\half\trd_{(\mu}H_{\nu)}{}^{\rho\sigma}+\half\trd^{(\rho}H^{\sigma)}{}_{\mu\nu}=\half g^{\rho\kappa}g^{\sigma\lambda}\left(
\trd_{[\nu}H_{\lambda]\mu\kappa}-\trd_{[\mu}H_{\kappa]\nu\lambda}\right)\,.
\ee
~\\
~\\


\section*{SM-B: Covariance of ${\Deltab}$}
Here we demonstrate the full covariance of the differential operation $\Deltab T_{A_{1}A_{2}\cdots A_{\bs}}$~(\ref{DELTA}) under DFT-diffeomorphisms. For the fundamental fields of DFT, such as $P_{AB}, \brP_{AB}, \cH_{AB}$, and  $e^{-2d}$, the active form of the diffeomorphisms  $\delta_{\xi}$  is  given  by the generalized Lie derivative $\hcL_{\xi}$~(\ref{hcL}).   This induces the following transformation of the DFT-Christoffel connection~\cite{Jeon:2011cn} which involves the six-index projectors~(\ref{P6}):
\be
\delta_{\xi}\Gamma_{CAB}=\hcL_{\xi}\Gamma_{CAB}
-2\partial_{C}\partial_{[A}\xi_{B]}+2(\cP+\brcP)_{CAB}{}^{DEF}\partial_{D}\partial_{[E}\xi_{F]}\,.
\label{AnomalyG}
\ee
Consequently,  the semi-covariant derivative of a generic tensor density~(\ref{semicovD}) is  generally not  fully covariant.  The difference between the actual transformation $\delta_{\xi}$ and the generalized Lie derivative~$\hcL_{\xi}$ is given by:
\be
\big(\delta_{\xi}-\hcL_{\xi}\big)\big(\na_{C}T_{A_{1}\cdots A_{\bs}}\big)=
\dis{\sum_{i=1}^{\,\bs}2(\cP{+\brcP})_{CA_{i}}{}^{BDEF}
\partial_{D}\partial_{E}\xi_{F}\,T_{A_{1}\cdots A_{i-1} BA_{i+1}\cdots A_{\bs}}\,.}
\label{AnomalynaT}
\ee
By repeatedly applying this result  with care, we further obtain,
\be
\ba{rll}
\big(\delta_{\xi}-\hcL_{\xi}\big)\big(\na_{B}\na_{C}T_{A_{1}\cdots A_{\bs}}\big)&=&2 (\cP{+\brcP})_{BC}{}^{DEFG}
\partial_{E}\partial_{F}\xi_{G}\,\na_{D}T_{A_{1}\cdots A_{\bs}}\\
{}&{}&+
\dis{\sum_{i=1}^{\,\bs}}\left[\ba{l}
2(\cP{+\brcP})_{CA_{i}}{}^{DEFG}T_{A_{1}\cdots A_{i-1} DA_{i+1}\cdots A_{\bs}}\na_{B}\left(
\partial_{E}\partial_{F}\xi_{G}\right)\\
+
2 (\cP{+\brcP})_{CA_{i}}{}^{DEFG}
\partial_{E}\partial_{F}\xi_{G}\,\na_{B}T_{A_{1}\cdots A_{i-1} DA_{i+1}\cdots A_{\bs}}\\
+
2 (\cP{+\brcP})_{BA_{i}}{}^{DEFG}
\partial_{E}\partial_{F}\xi_{G}\,\na_{C}T_{A_{1}\cdots A_{i-1} DA_{i+1}\cdots A_{\bs}}\ea
\right]\,.
\ea
\label{Anomalyna2T}
\ee
Furthermore, the field strength of the connection $\fR_{ABCD}$~(\ref{FSR}) and the semi-covariant four-index curvature $S_{ABCD}$~(\ref{RiemannS}) transform as:
\be
\ba{rll}
\big(\delta_{\xi}-\hcL_{\xi}\big)\fR_{ABCD}&=&
-2\Gamma^{H}{}_{AB}\partial_{H}\partial_{[C}\xi_{D]}
+2\Gamma^{E}{}_{CD}(\cP{+\brcP})_{EAB}{}^{HIJ}\partial_{H}\partial_{[I}\xi_{J]}+4\na_{[C}\Big((\cP{+\brcP})_{D]AB}{}^{HIJ}\partial_{H}\partial_{[I}\xi_{J]}\Big)\,,\\
\big(\delta_{\xi}-\hcL_{\xi}\big)S_{ABCD}&=&
2\na_{[A}\Big(\!(\cP{+\brcP})_{B][CD]}{}^{EFG}\partial_{E}\partial_{F}\xi_{G}\Big)
+2\na_{[C}\Big(\!(\cP{+\brcP})_{D][AB]}{}^{EFG}\partial_{E}\partial_{F}\xi_{G}\Big)\,.
\ea
\label{AnomalyfRS}
\ee
As a result, we  obtain:
\be
\scalebox{0.96}{$
\big(\delta_{\xi}-\hcL_{\xi}\big)\big(\fR_{[AB]}-\half\Gamma^{CD}{}_{A}\Gamma_{CDB}\big)=2\Gamma^{CD}{}_{B}\partial_{C}\partial_{[D}\xi_{A]}
-2\Gamma^{CD}{}_{A}(\cP{+\brcP})_{CDB}{}^{EFG}\partial_{E}\partial_{[F}\xi_{G]}
-(\cP{+\brcP})^{C}{}_{AB}{}^{DEF}\na_{C}\big(\partial_{D}\partial_{[E}\xi_{F]}\big)\,.$}
\label{dhcLfRG}
\ee

We recall the differential operation~(\ref{DELTA}): 
\be
\ba{rll}
\Deltab T_{A_{1}A_{2}\cdots A_{\bs}}&=&
P^{BC}\na_{B}\na_{C}T_{A_{1}\cdots A_{\bs}}\\
{}&{}&+\dis{\sum_{i=1}^{\,\bs}~2P_{A_{i}}{}^{C}P_{B}{}^{D}\left(\fR_{[CD]}-\half\Gamma^{EF}{}_{C}\Gamma_{EFD}-\Gamma^{E}{}_{CD}\na_{E}\right)T_{A_{1}\cdots A_{i-1}}{}^{B}{}_{A_{i+1}\cdots A_{\bs}}}\\
&{}&+\dis{\sum_{i<j}~2\!\left(\ba{l}
P_{A_{i}}{}^{D}P_{B}{}^{E}\fR_{A_{j}CDE}+P_{A_{j}}{}^{D}P_{C}{}^{E}\fR_{A_{i}BDE}\\
-2P_{A_{i}}{}^{D}P_{B}{}^{E}P_{A_{j}}{}^{F}P_{C}{}^{G}S_{DEFG}\ea\right)T_{A_{1}\cdots A_{i-1}}{}^{B}{}_{A_{i+1}\cdots A_{j-1}}{}^{C}{}_{A_{j+1}\cdots A_{\bs}}}\,,
\ea
\label{SMDELTA}
\ee
which apparently decomposes into three parts: $P^{BC}\na_{B}\na_{C}T_{A_{1}\cdots A_{\bs}}$, a single-index sum $\sum_{i}$, and a double-index sum $\sum_{i<j}$.  Using the results of (\ref{AnomalyG}), (\ref{AnomalynaT}), (\ref{Anomalyna2T}), (\ref{AnomalyfRS}), and an identity like $\sum_{i}\sum_{j}~\cA_{ij}=\sum_{i}\,\cA_{ii}+\sum_{i<j}\,\cA_{ij}+\cA_{ji}$, we obtain the transformation of each part: 

For the first part:
\be
\scalebox{0.94}{$
\big(\delta_{\xi}-\hcL_{\xi}\big)\big(P^{BC}\na_{B}\na_{C}T_{A_{1}\cdots A_{\bs}}\big)
=\dis{\sum_{i=1}^{\,\bs}}~
2\cP^{E}{}_{A_{i}B}{}^{FGH}\Big[\na_{E}\!\left(
\partial_{F}\partial_{[G}\xi_{H]}\right)T_{A_{1}\cdots A_{i-1}}{}^{B}{}_{A_{i+1}\cdots A_{\bs}}+2
\partial_{F}\partial_{[G}\xi_{H]}\,\na_{E}T_{A_{1}\cdots A_{i-1}}{}^{B}{}_{A_{i+1}\cdots A_{\bs}}\Big]
\,,$}
\label{DELTAano0}
\ee
where the traceless property of the six-index project has been used, $P^{AB}\cP_{ABCDEF}=0$;  

For the single-sum part:
\be
\ba{l}
\big(\delta_{\xi}-\hcL_{\xi}\big)\left[\dis{\sum_{i=1}^{\,\bs}~2P_{A_{i}}{}^{C}P_{B}{}^{D}\big(\fR_{[CD]}-\half\Gamma^{EF}{}_{C}\Gamma_{EFD}-\Gamma^{E}{}_{CD}\na_{E}\big)T_{A_{1}\cdots A_{i-1}}{}^{B}{}_{A_{i+1}\cdots A_{\bs}}}\right]\\
=\dis{\sum_{i=1}^{\,\bs}\left[
\ba{l}
\Big(4P_{A_{i}}{}^{C}P_{B}{}^{D}\Gamma^{EF}{}_{D}\partial_{E}\partial_{[F}\xi_{C]}
-2\cP^{E}{}_{A_{i}B}{}^{FGH}\na_{E}\big(\partial_{F}\partial_{[G}\xi_{H]}\big)\Big)T_{A_{1}\cdots A_{i-1}}{}^{B}{}_{A_{i+1}\cdots A_{\bs}}\\
-4\Big(\cP^{E}{}_{A_{i}B}{}^{DEF}\partial_{D}\partial_{[E}\xi_{F]}-P_{A_{i}}{}^{C}P_{B}{}^{D}\partial^{E}\partial_{[C}\xi_{D]}\Big)\na_{E}T_{A_{1}\cdots A_{i-1}}{}^{B}{}_{A_{i+1}\cdots A_{\bs}}
\ea\right]}\\
-\dis{\sum_{i<j}~4\Big[P_{A_{i}}{}^{D}P_{B}{}^{E}\Gamma^{F}{}_{DE}(\cP{+\brcP})_{FA_{j}C}{}^{HIJ}+P_{A_{j}}{}^{D}P_{C}{}^{E}\Gamma^{F}{}_{DE}(\cP{+\brcP})_{FA_{i}B}{}^{HIJ}\Big]
\partial_{H}\partial_{[I}\xi_{J]}} T_{A_{1}\cdots A_{i-1}}{}^{B}{}_{A_{i+1}\cdots A_{j-1}}{}^{C}{}_{A_{j+1}\cdots A_{\bs}}\\
=\dis{\sum_{i=1}^{\,\bs}
-2\cP^{E}{}_{A_{i}B}{}^{FGH}\Big[\na_{E}\big(\partial_{F}\partial_{[G}\xi_{H]}\big)T_{A_{1}\cdots A_{i-1}}{}^{B}{}_{A_{i+1}\cdots A_{\bs}}
+2\partial_{F}\partial_{[G}\xi_{H]}\na_{E}T_{A_{1}\cdots A_{i-1}}{}^{B}{}_{A_{i+1}\cdots A_{\bs}}\Big]}\\
\quad+\dis{\sum_{i<j}~4\left[\ba{l}P_{A_{i}}{}^{D}P_{B}{}^{E}\Big(
\Gamma^{E}{}_{A_{j}C}\partial_{E}\partial_{[D}\xi_{E]}-\Gamma^{F}{}_{DE}(\cP{+\brcP})_{FA_{j}C}{}^{GHI}\partial_{G}\partial_{[H}\xi_{I]}\Big)\\
+P_{A_{j}}{}^{D}P_{C}{}^{E}\Big(\Gamma^{E}{}_{A_{i}B}\partial_{E}\partial_{[D}\xi_{E]}-\Gamma^{F}{}_{DE}(\cP{+\brcP})_{FA_{i}B}{}^{GHI}\partial_{G}\partial_{[H}\xi_{I]}\Big)\ea\right]T_{A_{1}\cdots A_{i-1}}{}^{B}{}_{A_{i+1}\cdots A_{j-1}}{}^{C}{}_{A_{j+1}\cdots A_{\bs}}}\,,
\ea
\label{DELTAano1}
\ee
where the first equality holds due to (\ref{dhcLfRG}) while  the second one is valid  by  (\ref{PbrPG});

And for the double-sum part:
\be
\ba{l}
\big(\delta_{\xi}-\hcL_{\xi}\big)\left[\dis{\sum_{i<j}~2\!\left(\!\!\ba{l}
P_{A_{i}}{}^{D}P_{B}{}^{E}\fR_{A_{j}CDE}+P_{A_{j}}{}^{D}P_{C}{}^{E}\fR_{A_{i}BDE}\\
-2P_{A_{i}}{}^{D}P_{B}{}^{E}P_{A_{j}}{}^{F}P_{C}{}^{G}S_{DEFG}\ea\!\!\right)T_{A_{1}\cdots A_{i-1}}{}^{B}{}_{A_{i+1}\cdots A_{j-1}}{}^{C}{}_{A_{j+1}\cdots A_{\bs}}}\right]\\
=\dis{\sum_{i<j}~-4\!\left[\!\!\ba{l}
P_{A_{i}}{}^{D}P_{B}{}^{E}\Big(\Gamma^{F}{}_{A_{j}C}\partial_{F}\partial_{[D}\xi_{E]}-\Gamma^{F}{}_{DE}(\cP{+\brcP})_{FA_{j}C}{}^{GHI}\partial_{G}\partial_{[H}\xi_{I]}\Big)\\
+P_{A_{j}}{}^{D}P_{C}{}^{E}\Big(\Gamma^{F}{}_{A_{i}B}\partial_{F}\partial_{[D}\xi_{E]}-\Gamma^{F}{}_{DE}(\cP{+\brcP})_{FA_{i}B}{}^{GHI}\partial_{G}\partial_{[H}\xi_{I]}\Big)
\ea\!\!\right]T_{A_{1}\cdots A_{i-1}}{}^{B}{}_{A_{i+1}\cdots A_{j-1}}{}^{C}{}_{A_{j+1}\cdots A_{\bs}}\,.}
\ea
\label{DELTAano2}
\ee
It is straightforward to see that these three anomalous terms~(\ref{DELTAano0}), (\ref{DELTAano1}), and (\ref{DELTAano2}) add up to zero,  completing  the verification of the full covariance:
\be
\delta_{\xi}\left(\Deltab T_{A_{1}A_{2}\cdots A_{\bs}}\right)=\hcL_{\xi}\left(\Deltab  T_{A_{1}A_{2}\cdots A_{\bs}}\right)\,.
\ee
~\\
\vspace{23pt}
\vspace{23pt}

\section*{SM-C: Covariance of $\brDeltab$}
In a completely parallel manner, we can  verify the  full covariance of  the mirrored differential operation:
\be
\ba{rll}
\brDeltab T_{A_{1}A_{2}\cdots A_{\bs}}&:=&
\brP^{BC}\na_{B}\na_{C}T_{A_{1}A_{2}\cdots A_{\bs}}\\
{}&{}&+\dis{\sum_{i=1}^{\,\bs}~2\brP_{A_{i}}{}^{C}\brP_{B}{}^{D}\left(\fR_{[CD]}-\half\Gamma^{EF}{}_{C}\Gamma_{EFD}-\Gamma^{E}{}_{CD}\na_{E}\right)T_{A_{1}\cdots A_{i-1}}{}^{B}{}_{A_{i+1}\cdots A_{\bs}}}\\
&{}&+\dis{\sum_{i<j}~2\left(\ba{l}
\brP_{A_{i}}{}^{D}\brP_{B}{}^{E}\fR_{A_{j}CDE}+\brP_{A_{j}}{}^{D}\brP_{C}{}^{E}\fR_{A_{i}BDE}\\
-2\brP_{A_{i}}{}^{D}\brP_{B}{}^{E}\brP_{A_{j}}{}^{F}\brP_{C}{}^{G}S_{DEFG}\ea\right)T_{A_{1}\cdots A_{i-1}}{}^{B}{}_{A_{i+1}\cdots A_{j-1}}{}^{C}{}_{A_{j+1}\cdots A_{\bs}}}
\,.
\ea
\label{brDeltabFULL}
\ee
For each of the three parts, we obtain:
\be
\scalebox{0.94}{$
\big(\delta_{\xi}-\hcL_{\xi}\big)\big(\brP^{BC}\na_{B}\na_{C}T_{A_{1}\cdots A_{\bs}}\big)=\dis{\sum_{i=1}^{\,\bs}}~
2\brcP^{E}{}_{A_{i}B}{}^{FGH}\Big[\na_{E}\big(
\partial_{F}\partial_{[G}\xi_{H]}\big)
T_{A_{1}\cdots A_{i-1}}{}^{B}{}_{A_{i+1}\cdots A_{\bs}}+2\partial_{F}\partial_{[G}\xi_{H]}\,\na_{E}T_{A_{1}\cdots A_{i-1}}{}^{B}{}_{A_{i+1}\cdots A_{\bs}}\Big]\,,$}
\label{brDELTAano0}
\ee
\be
\ba{l}
\big(\delta_{\xi}-\hcL_{\xi}\big)\left[\dis{\sum_{i=1}^{\,\bs}~2\brP_{A_{i}}{}^{C}\brP_{B}{}^{D}\big(\fR_{[CD]}-\half\Gamma^{EF}{}_{C}\Gamma_{EFD}-\Gamma^{E}{}_{CD}\na_{E}\big)T_{A_{1}\cdots A_{i-1}}{}^{B}{}_{A_{i+1}\cdots A_{\bs}}}\right]\\
=\dis{\sum_{i=1}^{\,\bs}
-2\brcP^{E}{}_{A_{i}B}{}^{FGH}\Big[\na_{E}\big(\partial_{F}\partial_{[G}\xi_{H]}\big)T_{A_{1}\cdots A_{i-1}}{}^{B}{}_{A_{i+1}\cdots A_{\bs}}
+2\partial_{F}\partial_{[G}\xi_{H]}\na_{E}T_{A_{1}\cdots A_{i-1}}{}^{B}{}_{A_{i+1}\cdots A_{\bs}}\Big]}\\
\quad+\dis{\sum_{i<j}~4\left[\ba{l}\brP_{A_{i}}{}^{D}\brP_{B}{}^{E}\Big(
\Gamma^{E}{}_{A_{j}C}\partial_{E}\partial_{[D}\xi_{E]}-\Gamma^{F}{}_{DE}(\cP{+\brcP})_{FA_{j}C}{}^{GHI}\partial_{G}\partial_{[H}\xi_{I]}\Big)\\
+\brP_{A_{j}}{}^{D}\brP_{C}{}^{E}\Big(\Gamma^{E}{}_{A_{i}B}\partial_{E}\partial_{[D}\xi_{E]}-\Gamma^{F}{}_{DE}(\cP{+\brcP})_{FA_{i}B}{}^{GHI}\partial_{G}\partial_{[H}\xi_{I]}\Big)\ea\right]T_{A_{1}\cdots A_{i-1}}{}^{B}{}_{A_{i+1}\cdots A_{j-1}}{}^{C}{}_{A_{j+1}\cdots A_{\bs}}}\,,
\ea
\label{brDELTAano1}
\ee
and
\be
\ba{l}
\big(\delta_{\xi}-\hcL_{\xi}\big)\left[\dis{\sum_{i<j}~2\!\left(\!\!\ba{l}
\brP_{A_{i}}{}^{D}\brP_{B}{}^{E}\fR_{A_{j}CDE}+\brP_{A_{j}}{}^{D}\brP_{C}{}^{E}\fR_{A_{i}BDE}\\
-2\brP_{A_{i}}{}^{D}\brP_{B}{}^{E}\brP_{A_{j}}{}^{F}\brP_{C}{}^{G}S_{DEFG}\ea\!\!\right)T_{A_{1}\cdots A_{i-1}}{}^{B}{}_{A_{i+1}\cdots A_{j-1}}{}^{C}{}_{A_{j+1}\cdots A_{\bs}}}\right]\\
=\dis{\sum_{i<j}~-4\!\left[\!\!\ba{l}
\brP_{A_{i}}{}^{D}\brP_{B}{}^{E}\Big(\Gamma^{F}{}_{A_{j}C}\partial_{F}\partial_{[D}\xi_{E]}-\Gamma^{F}{}_{DE}(\cP{+\brcP})_{FA_{j}C}{}^{GHI}\partial_{G}\partial_{[H}\xi_{I]}\Big)\\
+\brP_{A_{j}}{}^{D}\brP_{C}{}^{E}\Big(\Gamma^{F}{}_{A_{i}B}\partial_{F}\partial_{[D}\xi_{E]}-\Gamma^{F}{}_{DE}(\cP{+\brcP})_{FA_{i}B}{}^{GHI}\partial_{G}\partial_{[H}\xi_{I]}\Big)
\ea\!\!\right]T_{A_{1}\cdots A_{i-1}}{}^{B}{}_{A_{i+1}\cdots A_{j-1}}{}^{C}{}_{A_{j+1}\cdots A_{\bs}}\,.}
\ea
\label{brDELTAano2}
\ee
Again, these three anomalous terms~(\ref{brDELTAano0}), (\ref{brDELTAano1}), and (\ref{brDELTAano2}) add up to zero, ensuring   full covariance. \\

\section*{SM-D: Computation of $\Deltab+\brDeltab=0$}
Here, we focus on  computing Equation~(\ref{SUM}), $\Deltab+\brDeltab=0$.  We  outline how the sum of the mirroring operators vanishes identically, subject to the section condition. 

We start with the following expression:
\be
\ba{rll}
\big(\Deltab+\brDeltab\big) T_{A_{1}A_{2}\cdots A_{\bs}}&=&
\na_{B}\na^{B}T_{A_{1}\cdots A_{\bs}}\\
{}&{}&+\dis{\sum_{i=1}^{\,\bs}~2\left(P_{A_{i}}{}^{C}P_{B}{}^{D}+\brP_{A_{i}}{}^{C}\brP_{B}{}^{D}\right)\left(\fR_{[CD]}-\half\Gamma^{EF}{}_{C}\Gamma_{EFD}-\Gamma^{E}{}_{CD}\na_{E}\right)T_{A_{1}\cdots A_{i-1}}{}^{B}{}_{A_{i+1}\cdots A_{\bs}}}\\
\multicolumn{3}{r}{
+\dis{\sum_{i<j}~2\!\left(\!\!\ba{l}
\left(P_{A_{i}}{}^{D}P_{B}{}^{E}+\brP_{A_{i}}{}^{D}\brP_{B}{}^{E}\right)\fR_{A_{j}CDE}\\
+\left(P_{A_{j}}{}^{F}P_{C}{}^{G}+\brP_{A_{j}}{}^{F}\brP_{C}{}^{G}\right)\fR_{A_{i}BFG}\\
-2\left(P_{A_{i}}{}^{D}P_{B}{}^{E}P_{A_{j}}{}^{F}P_{C}{}^{G}+\brP_{A_{i}}{}^{D}\brP_{B}{}^{E}\brP_{A_{j}}{}^{F}\brP_{C}{}^{G}\right)S_{DEFG}\ea\!\!\right)T_{A_{1}\cdots A_{i-1}}{}^{B}{}_{A_{i+1}\cdots A_{j-1}}{}^{C}{}_{A_{j+1}\cdots A_{\bs}}}
\,.}
\ea
\label{DbrDT}
\ee
To show this vanishes identically, we need to expand  the semi-covariant derivatives explicitly,
\be
\ba{rll}
\na_{B}\na^{B}T_{A_1A_2\cdots A_{\bs}}&=&
\dis{\sum_{i}~}\Big[\partial_{C}\Gamma^{C}{}_{A_{i}B}+(1-2\omega)\Gamma_{C}{}^{CD}\Gamma_{DA_{i}B}-\Gamma_{CDA_{i}}\Gamma^{CD}{}_{B}+2\Gamma^{C}{}_{A_{i}B}\partial_{C}\Big]T_{A_1\cdots A_{i-1}}{}^{B}{}_{A_{i+1}\cdots A_{\bs}}\\
{}&{}&+\dis{\sum_{i<j}~}2\Gamma_{DA_{i}B}\Gamma^{D}{}_{A_{j}C}\,T_{A_1\cdots A_{i-1}}{}^{B}{}_{A_{i+1}\cdots A_{j-1}}{}^{C}{}_{A_{j+1}\cdots A_{\bs}}\,,
\ea
\label{na2}
\ee
where the section condition has been consistently applied,  with  the relation $\Gamma^{B}{}_{BA}=-2\partial_{A}d$. Furthermore,  we note:
\be
\ba{l}
\dis{\sum_{i=1}^{\,\bs}~2\big(P_{A_{i}}{}^{C}P_{B}{}^{D}+\brP_{A_{i}}{}^{C}\brP_{B}{}^{D}\big)\Big(\fR_{[CD]}-\half\Gamma^{EF}{}_{C}\Gamma_{EFD}-\Gamma^{E}{}_{CD}\na_{E}\Big)T_{A_{1}\cdots A_{i-1}}{}^{B}{}_{A_{i+1}\cdots A_{\bs}}}\\
=\dis{\sum_{i=1}^{\,\bs}~-\big(P_{A_{i}}{}^{C}P_{B}{}^{D}+\brP_{A_{i}}{}^{C}\brP_{B}{}^{D}\big)
\Big(\partial_{E}\Gamma^{E}{}_{CD}+\Gamma_{E}{}^{EF}\Gamma_{FCD}
+\Gamma^{EF}{}_{C}\Gamma_{EFD}+2\Gamma^{E}{}_{CD}\na_{E}\Big)T_{A_{1}\cdots A_{i-1}}{}^{B}{}_{A_{i+1}\cdots A_{\bs}}}\\
=\dis{\sum_{i=1}^{\,\bs}~-\left[\!\ba{l}\scalebox{0.95}{$\big(P_{A_{i}}{}^{C}P_{B}{}^{D}+\brP_{A_{i}}{}^{C}\brP_{B}{}^{D}\big)
\Big(\partial_{E}\Gamma^{E}{}_{CD}+(1-2\omega)\Gamma_{E}{}^{EF}\Gamma_{FCD}
+\Gamma^{EF}{}_{C}\Gamma_{EFD}+2\Gamma^{E}{}_{CD}\partial_{E}\Big)$}\\
+2\big(P_{A_{i}}{}^{C}P_{F}{}^{D}+\brP_{A_{i}}{}^{C}\brP_{F}{}^{D}\big)
\Gamma_{ECD}\Gamma^{EF}{}_{B}
\ea\!
\right]T_{A_{1}\cdots A_{i-1}}{}^{B}{}_{A_{i+1}\cdots A_{\bs}}}\\
-\dis{\sum_{i<j}~}2\Big[\big(P_{A_{i}}{}^{E}P_{B}{}^{F}{+\brP}_{A_{i}}{}^{E}\brP_{B}{}^{F}\big)\Gamma_{DEF}\Gamma^{D}{}_{A_{j}C}+
\big(P_{A_{j}}{}^{E}P_{C}{}^{F}{+\brP}_{A_{j}}{}^{E}\brP_{C}{}^{F}\big)\Gamma_{DEF}\Gamma^{D}{}_{A_{i}B}\Big]T_{A_{1}\cdots A_{i-1}}{}^{B}{}_{A_{i+1}\cdots A_{j-1}}{}^{C}{}_{A_{j+1}\cdots A_{\bs}}\,,
\ea
\label{RCDna}
\ee
where the expression of $\fR_{[AB]}$~(\ref{fRsAB}) has been substituted.

The sum of the operators $\Deltab+\brDeltab$  now decomposes into two new distinct parts:   a single-index sum denoted by $\sum_{i}$ and   a double-index sum represented by  $\sum_{i<j}$.  Each  sum  vanishes separately.   The ``no sum'' part has already been trivialized  in (\ref{na2}).

\begin{itemize}
\item  \textbf{Single-index sum:} $\sum_{i}$.\\
To demonstrate that the single-index sum vanishes identically, we fully utilize the completeness of the projectors, $P_{A}{}^{C}+\brP_{A}{}^{C}=\delta_{A}{}^{C}$~(\ref{CO}), as well as the compatibility conditions, $\na_{C}P_{AB}=0=\na_{C}\brP_{AB}$~(\ref{compatibility}).  Additionally, we take into account that the free index  $E$  in the expression  $P_{A}{}^{C}\brP_{B}{}^{D}\Gamma^{E}{}_{CD}$ is  derivative-index-valued,  resulting  in (\ref{PbrPG}).  We  then obtain:
\be
\big(P_{A}{}^{C}P_{B}{}^{D}+\brP_{A}{}^{C}\brP_{B}{}^{D}\big)
\Big[(1-2\omega)\Gamma_{E}{}^{EF}\Gamma_{FCD}
+2\Gamma^{E}{}_{CD}\partial_{E}\Big]=(1-2\omega)\Gamma_{E}{}^{EF}\Gamma_{FAB}
+2\Gamma^{E}{}_{AB}\partial_{E}\,,
\label{S1}
\ee
\be
\ba{l}
\Gamma_{EFA}\Gamma^{EF}{}_{B}+\big(P_{A}{}^{C}P_{B}{}^{D}+\brP_{A}{}^{C}\brP_{B}{}^{D}\big)\Gamma^{EF}{}_{C}\Gamma_{EFD}+2\big(P_{A}{}^{C}P_{F}{}^{D}+\brP_{A}{}^{C}\brP_{F}{}^{D}\big)\Gamma_{ECD}\Gamma^{EF}{}_{B}\\
=\Big[\big(\brP_{A}{}^{C}P_{B}{}^{D}-P_{A}{}^{C}\brP_{B}{}^{D}\big)\cH^{FG}+2P_{A}{}^{C}P_{B}{}^{D}\brP^{FG}+2\brP_{A}{}^{C}\brP_{B}{}^{D}P^{FG}\Big]\Gamma_{EFC}\Gamma^{E}{}_{GD}\\
=\big(\brP_{A}{}^{C}P_{B}{}^{D}-P_{A}{}^{C}\brP_{B}{}^{D}\big)\cH^{FG}\Gamma_{EFC}\Gamma^{E}{}_{GD}\,,
\ea
\label{S2}
\ee
and,  as a rather nontrivial relation, 
\be
\ba{l}
\partial_{E}\Gamma^{E}{}_{AB}-\big(P_{A}{}^{C}P_{B}{}^{D}{+\brP}_{A}{}^{C}\brP_{B}{}^{D}\big)\partial_{E}\Gamma^{E}{}_{CD}\\
=\,\big(P_{A}{}^{C}\brP_{B}{}^{D}{+P}_{A}{}^{C}\brP_{B}{}^{D}\big)\partial_{E}\Gamma^{E}{}_{CD}\\
=\,-\Gamma^{E}{}_{CD}\partial_{E}\big(P_{A}{}^{C}\brP_{B}{}^{D}{+P}_{A}{}^{C}\brP_{B}{}^{D}\big)\\
=\,\Gamma^{ECD}\left[\ba{l}\Gamma_{EAF}\big(P^{F}{}_{C}\brP_{BD}{+\brP}^{F}{}_{C}P_{BD}\big)+\Gamma_{ECF}\big(P_{A}{}^{F}\brP_{BD}{+\brP}_{A}{}^{F}P_{BD}\big)\\
+\Gamma_{EBF}\big(P_{AC}\brP^{F}{}_{D}{+\brP}_{AC}P^{F}{}_{D}\big)
+\Gamma_{EDF}\big(P_{AC}\brP_{B}{}^{F}{+\brP}_{AC}P_{B}{}^{F}\big)\ea\right]\\
=\,-\Gamma^{E}{}_{CD}\Gamma_{EFA}\big(P^{CF}\brP^{D}{}_{B}+\brP^{CF}P^{D}{}_{B}\big)+\Gamma^{EC}{}_{D}\Gamma_{ECF}\big(P^{F}{}_{A}\brP^{D}{}_{B}+\brP^{F}{}_{A}P^{D}{}_{B}\big)\\
\quad~+\Gamma^{E}{}_{DC}\Gamma_{EFB}\big(\brP^{DF}P^{C}{}_{A}+P^{DF}\brP^{C}{}_{A}\big)
-\Gamma^{ED}{}_{C}\Gamma_{EDF}\big(P^{C}{}_{A}\brP^{F}{}_{B}+\brP^{C}{}_{A}P^{F}{}_{B}\big)\\
=\,\Gamma^{E}{}_{CD}\Big[\Gamma_{EFB}\big(\brP^{CF}P^{D}{}_{A}+P^{CF}\brP^{D}{}_{A}\big)-\Gamma_{EFA}\big(P^{CF}\brP^{D}{}_{B}+\brP^{CF}P^{D}{}_{B}\big)\Big]\\
=\,\Gamma^{E}{}_{CD}\Gamma_{EFG}\cH^{CF}\big(\brP^{D}{}_{A}P^{G}{}_{B}-P^{D}{}_{A}\brP^{G}{}_{B}\big)\,.
\ea
\label{S3}
\ee
By substituting (\ref{S1}), (\ref{S2}), and (\ref{S3}) into (\ref{na2}) and (\ref{RCDna}),  it becomes straightforward to see that the single-index sum vanishes identically. \\

\item \textbf{Double-index sum:} $\sum_{i<j}$. \\
To demonstrate that the double-index sum vanishes separately,   we utilize the  properties of  $\fR_{ABCD}$ and $S_{ABCD}$~(\ref{RSp}).  
With the expression of $S_{ABCD}$ in terms of $\fR_{ABCD}$ and $\Gamma_{CAB}$,  as described in (\ref{RiemannS}),  it follows that:
\be
\ba{l}
\scalebox{0.95}{$\big(P_{A_{i}}{}^{D}P_{B}{}^{E}{+\brP}_{A_{i}}{}^{D}\brP_{B}{}^{E}\big)\fR_{A_{j}CDE}
+\big(P_{A_{j}}{}^{F}P_{C}{}^{G}{+\brP}_{A_{j}}{}^{F}\brP_{C}{}^{G}\big)\fR_{A_{i}BFG}
-2\big(P_{A_{i}}{}^{D}P_{B}{}^{E}P_{A_{j}}{}^{F}P_{C}{}^{G}{+\brP}_{A_{i}}{}^{D}\brP_{B}{}^{E}\brP_{A_{j}}{}^{F}\brP_{C}{}^{G}\big)S_{DEFG}$}\\
=\big(P_{A_{i}}{}^{D}P_{B}{}^{E}{+\brP}_{A_{i}}{}^{D}\brP_{B}{}^{E}\big)\big(P_{A_{j}}{}^{F}P_{C}{}^{G}{+\brP}_{A_{j}}{}^{F}\brP_{C}{}^{G}\big)\big(\fR_{DEFG}+\fR_{FGDE}-2S_{DEFG}\big)\\
=\big(P_{A_{i}}{}^{D}P_{B}{}^{E}{+\brP}_{A_{i}}{}^{D}\brP_{B}{}^{E}\big)\big(P_{A_{j}}{}^{F}P_{C}{}^{G}{+\brP}_{A_{j}}{}^{F}\brP_{C}{}^{G}\big)\Gamma_{HDE}\Gamma^{H}{}_{FG}\,.
\ea
\label{fR2S}
\ee
Additionally,  with (\ref{PbrPG}), we have:
\be
\ba{l}
\Gamma_{HA_{i}B}\Gamma^{H}{}_{A_{j}C}-\Big[\big(P_{A_{i}}{}^{E}P_{B}{}^{F}{+\brP}_{A_{i}}{}^{E}\brP_{B}{}^{F}\big)\Gamma_{HEF}\Gamma^{H}{}_{A_{j}C}+
\big(P_{A_{j}}{}^{E}P_{C}{}^{F}{+\brP}_{A_{j}}{}^{E}\brP_{C}{}^{F}\big)\Gamma_{HEF}\Gamma^{H}{}_{A_{i}B}\Big]\\
=\Big[\delta_{A_{i}}{}^{D}\delta_{B}{}^{E}\delta_{A_{j}}{}^{F}\delta_{C}{}^{G}-
\big(P_{A_{i}}{}^{D}P_{B}{}^{E}+\brP_{A_{i}}{}^{D}\brP_{B}{}^{E}\big)\delta_{A_{j}}{}^{F}\delta_{C}{}^{G}-\delta_{A_{i}}{}^{D}\delta_{B}{}^{E}
\big(P_{A_{j}}{}^{F}P_{C}{}^{G}+\brP_{A_{j}}{}^{F}\brP_{C}{}^{G}\big)\Big]\Gamma_{HDE}\Gamma^{H}{}_{FG}\\
=\Big[\big(P_{A_{i}}{}^{D}\brP_{B}{}^{E}{+\brP}_{A_{i}}{}^{D}P_{B}{}^{E}\big)
\big(P_{A_{j}}{}^{F}\brP_{C}{}^{G}{+\brP}_{A_{j}}{}^{F}P_{C}{}^{G}\big)-\big(P_{A_{i}}{}^{D}P_{B}{}^{E}{+\brP}_{A_{i}}{}^{D}\brP_{B}{}^{E}\big)
\big(P_{A_{j}}{}^{F}P_{C}{}^{G}{+\brP}_{A_{j}}{}^{F}\brP_{C}{}^{G}\big)
\Big]\Gamma_{HDE}\Gamma^{H}{}_{FG}\\
=-\big(P_{A_{i}}{}^{D}P_{B}{}^{E}{+\brP}_{A_{i}}{}^{D}\brP_{B}{}^{E}\big)
\big(P_{A_{j}}{}^{F}P_{C}{}^{G}{+\brP}_{A_{j}}{}^{F}\brP_{C}{}^{G}\big)\Gamma_{HDE}\Gamma^{H}{}_{FG}\,.
\ea
\label{GGD}
\ee
Substituting (\ref{fR2S}) and (\ref{GGD}) into (\ref{DbrDT}),  it becomes clear that the double-index sum vanishes identically.  This completes our verification of $\Deltab+\brDeltab=0$.\\
\end{itemize}

\end{widetext}

\end{document}